\documentclass{appolb}

\usepackage{graphicx}

\begin{document}

\title{From Quark-Gluon Plasma to the Perfect Liquid
\thanks{Lectures presented at the Cracow School of Theoretical Physics, XLVII Course, 
Zakopane, Poland, 14--22 June 2007}
}

\author{Berndt M\"uller
\address{Department of Physics, Duke University, Durham, NC 27708, USA}
}
\maketitle
\begin{abstract}
After reviewing some basic concepts of the theory of strongly interacting matter above normal nuclear energy density and reviewing some salient results of the experimental program at the Relativistic Heavy Ion Collider (RHIC), these lectures explain why the quark-gluon plasma observed in the RHIC experiments has been called a ``perfect liquid.'' They then give an introduction to some recent ideas concerning the possible origin of the nearly inviscid nature of the quark-gluon plasma and discuss the connection between low viscosity and strong parton energy loss of hot QCD matter.
\end{abstract}

\section{QCD matter under extreme conditions}

Nuclear matter can be compressed in two very different ways: A slow squeeze results in cold matter at high net baryon density; a rapid squeeze produces hot matter high in energy density. The first type of compression is impossible to perform in the laboratory, but it can be achieved in the cosmic environment by stellar collapse: The cores of old neutron stars are thought to contain nuclear matter at up to ten times normal baryon density and at rather low temperatures on the nuclear scale, in the keV range. Hot, dense nuclear matter can be produced in the laboratory by colliding two heavy nuclei at high energies. The higher the collision energy, the lower is the net baryon content of the matter formed in the center of mass. Of course, hot nuclear matter also existed for a brief time shortly after the Big Bang, with temperatures exceeding 200 MeV (about $2\times 10^{12}$ K) for the first 10 $\mu$s in the history of our universe.

Nuclear collisions of this kind occur regularly in cosmic ray reactions, but these are rare and difficult to study. Systematic studies using high-energy accelerators now have a history of over 30 years, beginning with the BEVALAC at Berkeley, followed by the BNL-AGS and the CERN-SPS, and culminating presently at the Relativistic Heavy Ion Collider (RHIC) at BNL. In the near future, even higher energies will be explored at the CERN-LHC, and large net baryon densities will be studied at RHIC and, later, at GSI-FAIR. The goal of all these experimental programs is to produce and study equilibrated nuclear matter at energy densities far exceeding that of ground state nuclear matter ($\varepsilon_0 \approx 0.14$ GeV/fm$^3$). As we shall discuss below, this goal has been achieved with spectacular results, especially at RHIC.

The theory of nuclear matter under extreme conditions is based on the fundamental theory of strongly interacting matter, quantum chromodynamics (QCD). This gauge theory of SU(3)-color describes the interactions among quarks and gluon by means of the Lagrangian
\begin{equation}
L_{\rm QCD} = - \frac{1}{4} G^a_{\mu\nu} G^{a\mu\nu}
  + \sum_f \bar{\psi}_f \gamma^\mu \left(i\partial_\mu - g A^a_\mu t^a \right) \psi_f
  + \sum_f m_f \bar{\psi}_f \psi_f .
\label{eq:LQCD}
\end{equation}
In this expression, $A^a_\mu, G^a_{\mu\nu}$ denote the gluon potential and field strength tensor, respectively, with the color index $a$ running from 1 to $N_c^2-1=8$. $\psi_f$ denotes the quark field of flavor $f = u, d, s, \ldots$, $\gamma^\mu$ are the Dirac matrices, $t^a$ the generators of SU(3) in the fundamental representation, and $m_f$ the (current) quark masses. 

At high density, the separation between quanta becomes small and either the Fermi momentum or the thermal momentum becomes large. The asymptotic freedom of QCD then suggests that interactions among the quarks and gluons are relatively weak, and their contribution to the energy density of the matter should be relatively small. We can then estimate the energy density as a function of the temperature $T$. Denoting the number of internal degrees of freedom by $\nu$ and assuming zero net baryon density ($\mu_B=0$) we have
\begin{equation}
\varepsilon \approx \nu \int \frac{d^3p}{(2\pi)^3} E \left( e^{E/T} \pm 1 \right)^{-1} 
\approx \left(\nu_G + \frac{7}{8}\nu_Q \right) \frac{\pi^2}{30} T^4 ,
\label{eq:eT}
\end{equation}
where we have neglected the quark masses in the last step. The degrees of gluons and quarks are easily counted: $\nu_G = 2(N_c^2-1)$ [spin and color] and $\nu_Q = 4 N_c N_f$ [spin, baryon number, color, and flavor]. For $N_f=2$ flavors ($u, d$) one finds that a temperature of 160 MeV is needed to reach an energy density of 1 GeV/fm$^3$, roughly 7.5 times that of normal nuclear matter. 

\begin{figure}[htb]
\centerline{\includegraphics[width=4.5in]{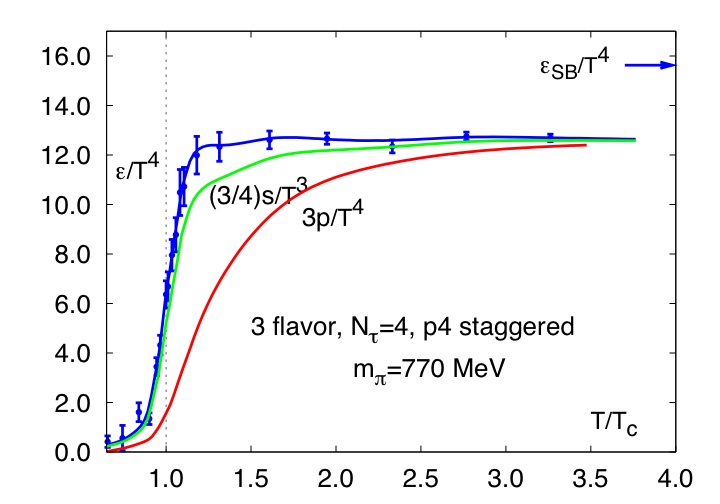}}
\caption{Thermodynamic quantities $\varepsilon, P, s$ divided by $T^4$ for QCD matter as function of temperature $T$ for $\mu_B=0$. The quantity $(45/2\pi^2)s/T^4$ can be considered as a measure of the number of thermodynamically active degrees of freedom.}
\label{fig:eos}
\end{figure}

If one wants to obtain more reliable information about the equation of state of QCD matter or wants to find out in which temperature domain the rough estimate (\ref{eq:eT}) applies, one must turn to exact calculations of the energy density. At present, this requires massive computer simulations of QCD discretized on a lattice. Over the past few years, increasingly precise lattice calculations of thermal QCD, extrapolated to the continuum and thermodynamic limits and to small quark masses, have become available (see Fig.~\ref{fig:eos}). These show a rapid transition from a low-energy region to a region with nearly constant $\varepsilon/T^4$ at $T_c \approx 160$ MeV. Detailed studies of the baryon number susceptibilities have shown that below $T_c$ the matter can be understood as a rather dilute gas of hadrons. Above $T_c$, the matter behaves as a SU(3)-colored plasma of interacting quarks and gluons. The various susceptibilities (specific heat, scalar quark density, Polyakov loop) exhibit strong peaks at $T_c$, but they do not diverge. The transition between the low-temperature and the high-temperature domains is thus not a true phase transition, but has the character of a rapid cross-over \cite{Brown:1990ev}. Precisely how wide the cross-over region is and whether it is characterized by a universal value of $T_c$ or slightly different values for different physical observables, is a matter of intense discussion \cite{Aoki:2006br,Cheng:2006qk}.

\section{The quark-gluon plasma}

\begin{figure}[htb]
\centerline{\includegraphics[width=3.5in]{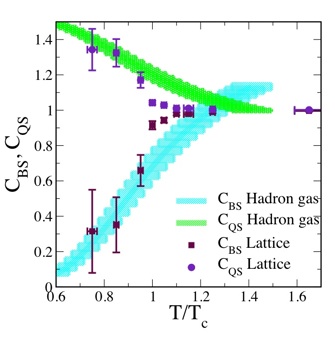}}
\caption{Flavor susceptibility ratios $C_{BS}$ and $C_{QS}$ as a function of $T/T_c$. The results of lattice simulations \protect\cite{Gavai:2005yk} are compared with expectations for a weakly coupled hadronic gas and a quasi-particulate quark-gluon plasma \protect\cite{Majumder:2005jy}. The transition to $C_{BS} = C_{QS} = 1$ expected for matter in which baryon number is carried by quarks occurs in a narrow range around $T_c$.}
\label{fig:CBS-CQS}
\end{figure}

What is clear from the lattice results is that above $T_c$ many quantities are well described by positing that the thermodynamically active constituents of QCD matter have the quantum number of quarks, not of hadrons. A striking example are the off-diagonal flavor susceptibilities. If a quasi-particle picture applies to the quark-gluon plasma, the ratio \cite{Koch:2005vg}
\begin{equation}
C_{XY} = 3 \frac{\langle XY \rangle - \langle X \rangle \langle Y \rangle}
                             {\langle Y^2 \rangle - \langle Y \rangle^2}
\label{eq:CXY}
\end{equation} 
of the covariance of the flavor quantum numbers $X$ and $Y$ and the variance of $Y$ alone is a sensitive indicator of the flavor quantum numbers of the plasma constituents \cite{Majumder:2006nq}. In the narrow temperature range $T = (1\pm 0.1)T_c$ lattice calculations show a striking transition of the quantities $C_{BS}$ and $C_{QS}$ from the values expected for a hadron gas containing strange baryons with integer electric charge $Q$ and baryon number $B$ to matter whose carriers of baryon number and electric charge have the fractional quantum numbers characteristic of strange quarks (see Fig.~\ref{fig:CBS-CQS}).

\begin{figure}[htb]
\centerline{\includegraphics[width=4.5in]{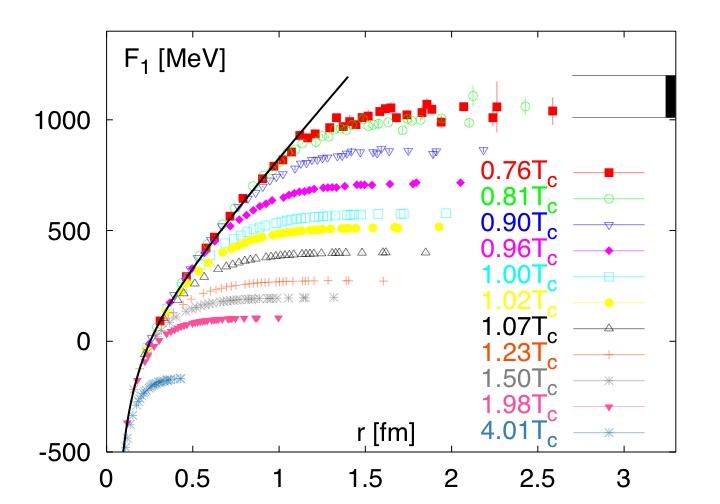}}
\caption{Free energy $F_1$ of a pair of heavy quarks forming a color singlet as a function of $Q\bar{Q}$ separation $r$ for various different temperatures. The potential is increasingly screened as the temperature increases. The screening effect at $T<T_c$ is due to pair production of light quarks from the vacuum when $R$ is large.}
\label{fig:QQpotential}
\end{figure}

How high in temperature one needs to go for gluons to also be described in such a picture is still unclear, but the thermodynamic functions are reproduced well in an approach based on perturbative quasi-particles for $T \ge 3 T_c$ \cite{Andersen:2002ey,Blaizot:2003iq}. These gluonic quasi-particle modes acquire a thermal ``mass'' of order $gT$, which helps avoid many infrared problems of the bare thermal gauge theory. An immediate consequence is the screening of the long-range color force responsible for quark confinement at zero temperature. In the weak coupling limit, the static color-electric potential $A_0^a$ of a heavy color charge embedded in the quark-gluon plasma satisfies the Poisson equation
\begin{equation}
- \nabla^2 A_0^a = g \rho_G^a(A_0^a) + g \rho_Q^a(A_0^a) ,
\label{eq:Poisson}
\end{equation}
where $\rho_{G/Q}^a(A_0^a) = - \mu_{G/Q}^2 A_0^a$ are the induced color polarization densities of gluons and quarks in the plasma, respectively. The perturbative calculation yields
\begin{equation}
\mu_{G}^2 = (gT)^2 ; \qquad \mu_{Q}^2 = N_f(gT)^2/6 ,
\label{eq:mu2}
\end{equation}
implying an exponentially screened static color potential with a range $\mu_{G/Q}^{-1} < 0.5$ fm. This picture suggested by a perturbative treatment is borne out by lattice simulations of the free energy of a pair of heavy, static quarks \cite{Kaczmarek:2005ui} (see Fig.~\ref{fig:QQpotential}).

\begin{figure}[htb]
\centerline{\includegraphics[width=4.5in]{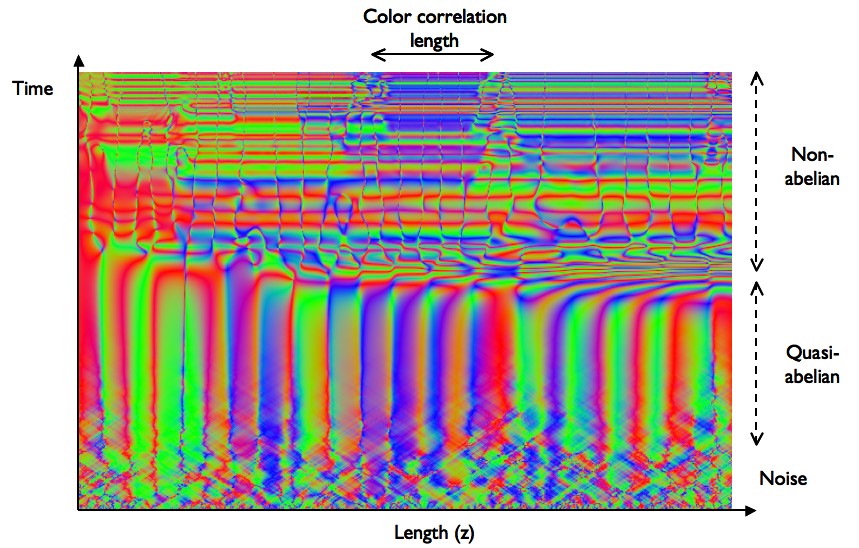}}
\caption{Spatial and temporal structure of dynamically generated color fields in a spatially one-dimensional quark-gluon plasma with an anisotropic momentum distribution \protect\cite{Strickland:2005we}.}
\label{fig:domains}
\end{figure}

For a long time it was thought that extended, though screened, color fields could not be generated in a quark-gluon plasma, because there exists no coherent source for such fields. (Owing to the absence of free quarks in nature, one cannot construct a ``color battery.'') Over the past decade, however, it has been realized that dynamic plasma mechanisms permit the formation of extended color fields by a kind of instability driven ``color dynamo.''\cite{Mrowczynski:1993qm,Mrowczynski:1994xv} These instabilities are generalizations of the phenomena first identified by Weibel in electromagnetic plasmas with an anisotropic momentum distribution \cite{Weibel:1959}. Fluctuations in the (color) current densities of such a plasma are found to induce (color-)magnetic fields which, in turn, amplify the original fluctuations producing a run-away effect. Momentum anisotropies are expected in a heavy-ion collision, where the longitudinal motion along the beam axis provides for a natural source of momentum space anisotropy. In a quark-gluon plasma, the exponentially growing fields eventually saturate due to the nonlinear interactions present in the Yang-Mills equation governing the SU(3) gauge field \cite{Arnold:2005vb}. Detailed simulations (see Fig.~\ref{fig:domains}) have shown that the fields in the steady state are characterized by an intensity of order $\langle E^2 \rangle \approx \langle B^2 \rangle \sim g^2 T^4$ and a correlation length of order $(gT)^{-1}$ \cite{Rebhan:2005re}.

\begin{figure}[htb]
\centerline{\includegraphics[width=3.6in]{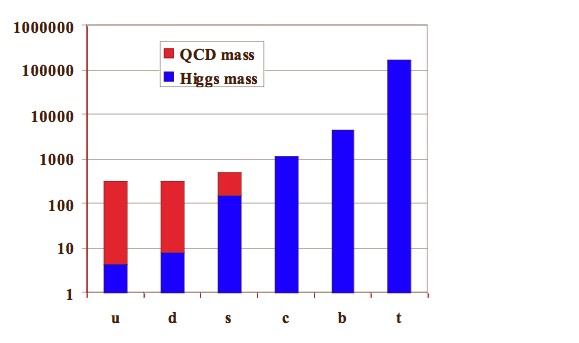}
\hspace{-0.6in} \includegraphics[width=1.8in]{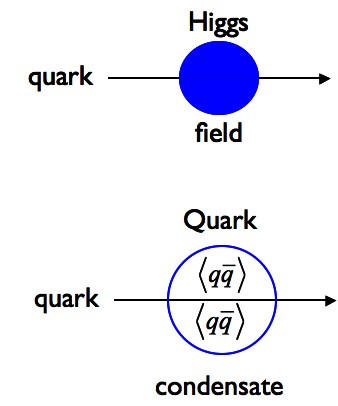}}
\caption{Right panel: The overwhelming part of the mass of ``constituent'' $u, d$, and $s$ quarks in hadrons in the normal QCD vacuum are dynamically generated by the vacuum quark condensate (bottom). This contribution to the mass disappears above $T_c$ and only the current mass generated by the Higgs field (top) remains. Left panel: Current mass due to Higgs field (blue) and constituent mass due to quark condensate (red) of the six known quark flavors. The quark mass is shown in MeV on a logarithmic scale.}
\label{fig:mass}
\end{figure}

A second conceptually important difference between normal nuclear or hadronic matter and the quark-gluon plasma phase is the restoration of the spontaneously broken chiral symmetry due to the disappearance of the vacuum quark condensate $\langle 0 | \bar{q} q | 0 \rangle$ at high temperature. As a result, the dynamical (``constituent'') masses of the light quarks $u, d$ ($s$) of order 300 MeV (500 MeV) give way to the much smaller current masses of order $5-10$ MeV (100 MeV) believed to be generated by the interaction with the Higgs field, as illustrated in Fig.~\ref{fig:mass}. It should be noted, however, that the interaction of the quarks with the thermal medium induces a new type of dynamical mass of order $gT$, which does not violate chiral symmetry. Although the phenomenon of chiral symmetry restoration is well established owing to lattice simulations, it is not easy to identify experimentally accessible signatures of this phenomenon. Electromagnetic probes, such as lepton pairs, may provide the best handle, but it remains uncertain how the effects of quark deconfinement and chiral symmetry restoration in the emission spectrum can be separated.

Rigorous algorithms for the simulation of lattice QCD only exist for zero net baryon density. In recent years, however, several ingenious techniques have been invented that allow to extrapolate the simulations to nonzero values of the baryon chemical potential $\mu_B$ \cite{Fodor:2001pe,Fodor:2002km,deForcrand:2002ci,Allton:2002zi}. Most of these simulations predict that the cross-over regime between hadronic matter and quark-gluon plasma narrows as $\mu_B$ increases and eventually gives way to a discontinuous phase transition. The critical endpoint of the first-order transition line may lie as low as $\mu_B \approx 300$ MeV \cite{Fodor:2004nz}. Unfortunately, the precise location of the critical point has been found to be extremely sensitive to the values of the current quark masses. Some calculations even indicate that the critical point might not exist \cite{deForcrand:2006pv}. Following the phase boundary to even higher values of $\mu_B$ one may encounter a triple point, where the line separating hadronic matter from the quark-gluon plasma meets the phase boundary between quark-gluon plasma and color superconducting quark matter \cite{Fodor:2007vv}. It must be noted, however, that such studies are still very much exploratory and speculative.

\section{Results from RHIC}

The Relativistic Heavy Ion Collider (RHIC) at Brookhaven National Laboratory was contructed to heat nuclear matter beyond $T_c$ by colliding heavy nuclei at sufficiently high energy, and thus to discover the quark-gluon plasma and explore its properties \cite{Harris:1996zx}. Because lattice QCD simulations are not yet able to reliably address the dynamic processes that govern the plasma evolution and response to perturbations, Initial expectations for the dynamic properties of the quark-gluon plasma were based on approximate analytical QCD calculations which assume that the color forces between quarks and gluons at high temperature are weak and can be treated perturbatively. Such analytical calculations suggested that the quark-gluon plasma should behave like a dilute gas, a loose assembly of particles which explodes in a more or less spherical pattern. The RHIC experiments have provided spectacular evidence that the quark-gluon plasma defies this expectation, at least in the temperature domain $T \leq 2T_c$  accessible at RHIC. Far from behaving like a dilute gas, the matter created in collisions of heavy nuclei flows like a nearly ``perfect'' liquid of liberated quarks, with almost undetectable viscosity.  

	The insights into the nature of quark-gluon plasma have made the first phase of RHIC a great success. In its first six runs (2000-2006), RHIC has provided four different collisions systems (Au+Au, d+Au, Cu+Cu, and p+p) at a variety of energies, ranging from nucleon-nucleon center-of-mass energies of 19.6 to 200 GeV.  The largest data samples were collected at the highest energy of 200 GeV, where the accelerator has achieved sustained operation at four times the design luminosity. The ability to study proton-proton, deuteron-nucleus, and nucleus-nucleus collisions at identical center-of-mass energies with the same detectors has been key to systematic control of the measurements. The experiments also measure the impact parameter (distance of closest approach) and the orientation of the reaction plane in each event. This detailed categorization of the collision geometry provides a wealth of differential observables, which have proven to be essential for precise, quantitative study of the plasma. The results obtained by the four RHIC experiments (BRAHMS, PHENIX, PHOBOS, and STAR), published in nearly 200 journal articles and four summary publications \cite{Arsene:2004fa,Back:2004je,Adams:2005dq,Adcox:2004mh}, are in remarkable quantitative agreement with each other. 

\begin{figure}[htb]
\centerline{\includegraphics[width=4.0in]{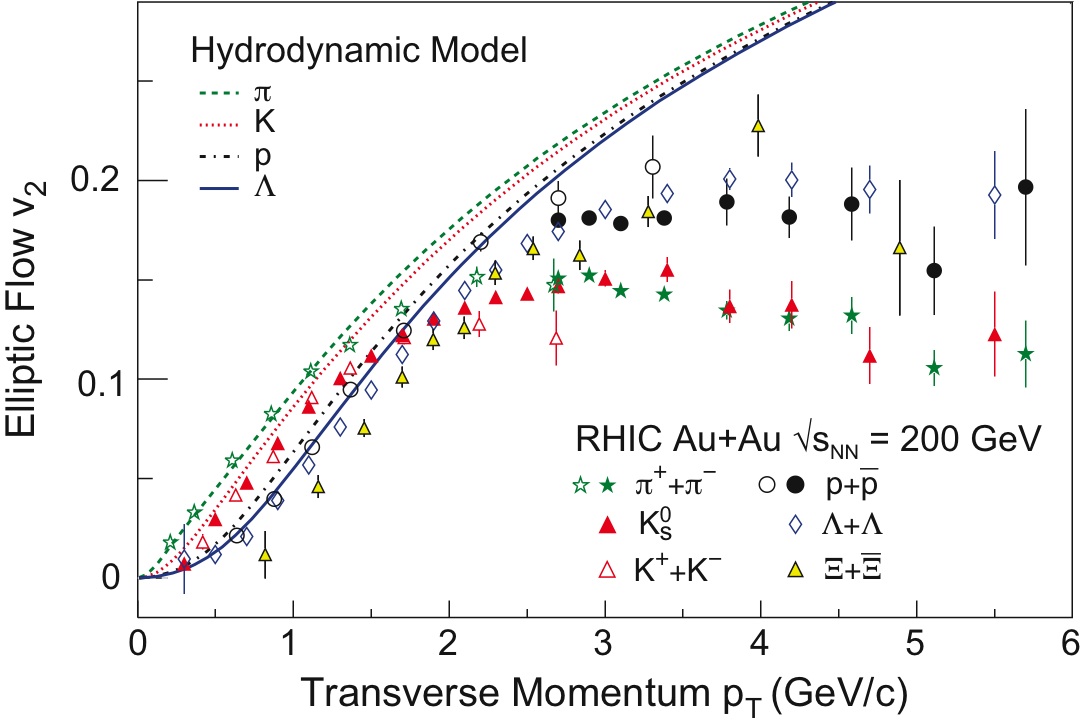}}
\caption{The momentum dependence of the flow anisotropy $v_2$ (``elliptic flow'') of hadrons measured in on-central Au+Au collisions at RHIC shows the characteristic dependence on hadron mass predicted by ideal relativistic hydrodynamics.}
\label{fig:v2}
\end{figure}

The initial set of heavy ion results from RHIC has provided evidence for the creation of a new state of thermalized matter at an unprecedented energy density of $(30-100)\varepsilon_0$, which exhibits almost ideal hydrodynamic behavior. Important results from the RHIC experiments include \cite{Muller:2006ee}:
\begin{itemize}
\setlength{\itemsep}{0pt}
\item chemical (flavor) and thermal equilibration of all observed hadrons including multi-strange baryons; only the reaction $gg \to \bar{s}s$ is known to achieve this on the time-scale of the nuclear reaction;
\item strong elliptic flow, indicating early thermalization (at times less than 1 fm/c) and a very low viscosity of the produced medium;
\item collective flow patterns related to independently flowing valence quarks, not hadrons;
\item strong jet quenching, implying a very large parton energy loss in the medium and a high color opacity of the produced matter;
\item strong suppression of open heavy flavor mesons at high transverse momentum, implying a large energy loss of heavy ($c$ and $b$) quarks in the medium;
\item direct photon emission at high transverse momentum that remains unaffected by the medium;
\item charmonium suppression effects that are similar to those observed at the lower energies of the CERN-SPS.
\end{itemize}
Among the results, three discoveries are most relevant to this lecture:

{\it ``Elliptic'' flow:}  The measured hadron spectra and their angular distributions in non-central heavy-ion collisons reveal the enormous collective motion of the medium.  In addition, measurements of electrons from the decays of hadrons containing charm quarks indicate that even heavy quarks flow with the bulk medium. These conclusions are based on measurements of the second Fourier coefficient $v_2$ of the azimuthal distribution of particles around the beam axis \cite{Ollitrault:1992bk,Poskanzer:1998yz}:
\begin{equation}
\frac{d^2N}{p_Tdp_Td\phi_p} = \frac{dN}{2\pi p_Tdp_T} \left( 1 + 2 v_2(p_T) \cos 2\phi_p + \cdots \right) ,
\label{eq:v2}
\end{equation}
where the angle $\phi_p$ is measured with respect to the reaction plane. The RHIC data are in good agreement with predictions for the hydrodynamic expansion of a nearly viscosity-free liquid -- often referred to as a ``perfect'' liquid -- from an initial oval-shaped configuration characterized by an anisotropic pressure gradient. The relative abundances of the produced hadrons with transverse momenta below 1.5 GeV/c, the shapes of their transverse momentum spectra, and their anisotropic or elliptic flow patterns are well described by ideal relativistic hydrodynamics with an equation of state similar to the one predicted by lattice QCD. Ideal hydrodynamics predicts a systematic Òfine structureÓ in the behavior of the elliptic flow strength parameter $v_2$ as a function of transverse momentum $p_T$ for hadrons of different mass  (see Fig.~\ref{fig:v2}). The magnitude of the observed collective flow points to rapid thermalization and equilibration of the matter on a time-scale of less than 1 fm/c \cite{Heinz:2001xi}. 

In reality, hydrodynamics is never ideal; any (non-super-)fluid has non-vanishing viscosities $\eta, \zeta$, which describe the deviation of the stress-energy tensor
\begin{equation}
T^{\mu\nu} = \varepsilon u^\mu u^nu - P \Delta^{\mu\nu} 
  + \eta \left(\nabla^\mu u^\nu + \nabla^\nu u^\mu - \frac{2}{3} \delta^{\mu\nu} \nabla\cdot u \right)
  + \zeta \Delta^{\mu\nu} \nabla\cdot u 
\label{eq:Tmunu}
\end{equation}
from its ideal form in the presence of flow gradients. The expression (\ref{eq:Tmunu}) is valid in the so-called energy frame and makes use of the abbreviations $\Delta^{\mu\nu} = g^{\mu\nu}-u^\mu u^\nu$ and $\nabla^\mu = \Delta^{\mu\nu} \partial_\nu$. $\eta$ is called the shear viscosity, $\zeta$ the bulk viscosity. $\zeta$ vanishes for a scale invariant fluid and is thus expected to be small for QCD matter, except possibly in the immediate vicinity of $T_c$, because scale invariance is only broken in QCD by the small current quark masses and by quantum effects. One of the exciting theoretical discoveries of the past few years is the insight that there may be a lower bound on the ratio between the shear viscosity $\eta$ and entropy density $s$ of any fluid: $4\pi\eta/s \geq 1$ \cite{Kovtun:2004de}. A ``perfect'' liquid is a fluid that attains this lower bound. There is mounting evidence from analysis of RHIC data that the matter produced is nearly such a perfect liquid, with a viscosity to entropy density ratio not larger than a factor of four times the lower bound. 

{\it Quark recombination:} Evidence that the medium is composed of deconfined, thermalized and collectively flowing quarks comes from detailed measurements of the spectra of a wide variety of hadrons. Baryons, which contain three valence quarks, show yields that are strongly enhanced relative to those of mesons, containing a valence quark and anti-quark, at intermediate transverse momentum ($p_T \sim 2-5$ GeV/c) in nuclear collisions, compared with p+p collisions.  This observation is well-described by models in which baryons and mesons are generated by the recombination of quarks drawn from a collectively flowing, thermally equilibrated partonic medium. For transverse momenta $p_T$ large enough so that hadron masses can be neglected, the recombination model predicts hadron spectra to have the form \cite{Fries:2003vb,Fries:2003kq,Greco:2003xt,Greco:2003mm}:
\begin{eqnarray}
E \frac{d N_M}{d^3 p} & \sim & \int \frac{d^3p_1 d^3p_2}{(2\pi)^6}
   w_q(p_1) \, w_{\bar{q}}(p_2) \delta(\vec{p}-\vec{p}_1-\vec{p}_2) 
   \sim e^{-E/T} ;
\nonumber \\
E \frac{d N_B}{d^3 p} & \sim & \int \left( \prod_{i=1}^3 \frac{d^3p_i}{(2\pi)^3} w_q(p_i) \right) 
   \, \delta\left(\vec{p}-\sum_i \vec{p}_i\right) 
   \sim e^{-E/T} ;
\label{eq:reco}
\end{eqnarray}
for mesons (M) and baryons (B), respectively. Here $w_q (w_{\bar{q}})$ denotes the phase space density of quarks (antiquarks) in the medium just prior to hadronization. These expressions predict meson and baryon yields of approximately equal magnitude at the same transverse momentum $p_T$, precisely what is observed in Au+Au collisions at RHIC, where the $p/\pi^+$ and $\bar{p}/\pi^-$ ratios are found to be near unity at $p_T = 2-3$ GeV/c. In contrast, hadron production by quark or gluon fragmentation predicts baryon-to-meson ratios much smaller than unity.

\begin{figure}[htb]
\vspace{0.2in}
\centerline{\includegraphics[width=4.5in]{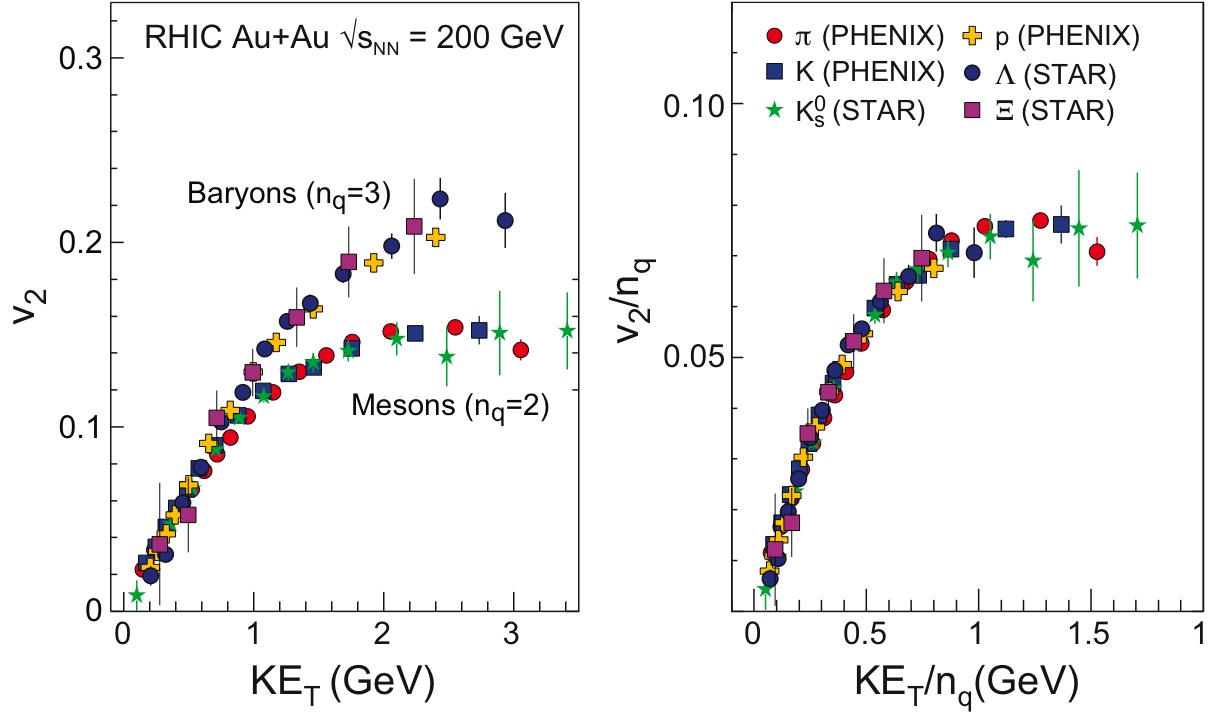}}
\caption{The transverse kinetic energy ($\tilde{E}_T = m_T-m$) dependences of $v_2$ of various mesons and baryons (left panel) collapse into a universal curve when $v_2$ and $\tilde{E}_T$ are scaled by the number of valence quarks of the hadron (right panel). This behavior is characteristic for hadron emission  by quark recombination from a collectively flowing, thermal quark-gluon plasma.}
\label{fig:v2scaled}
\end{figure}

The recombination model also suggests that the elliptic flow parameters of mesons and baryons are related to the elliptic flow parameter of quarks by the scaling law (assuming quarks and antiquarks exhibit the same collective flow) \cite{Fries:2003kq,Molnar:2003ff}:
\begin{equation}
v_2^{\rm (M)}(p_T) \approx 2\, v_2^{\rm (q)}(p_T/2) ; \qquad
v_2^{\rm (B)}(p_T) \approx 3\, v_2^{\rm (q)}(p_T/3) .
\label{eq:v2scal}
\end{equation}
Indeed, the measured elliptic flow patterns of baryons and mesons show remarkable agreement when scaled by the number of valence quarks (see Fig.~\ref{fig:v2scaled}). A combination of hydrodynamics and recombination model considerations has been used to suggest a slightly modified scaling behavior for $v_2$ when analyzed as a function of transverse kinetic energy $\tilde{E}_T = (m^2+p_T^2)^{1/2}-m$, resulting in two distinct branches, one for mesons and the other for baryons (Fig.~\ref{fig:v2scaled}, left panel). When both $v_2$ and $\tilde{E}_T$ are scaled by the number of valence quarks (2 for mesons, 3 for baryons) the two branches merge into a universal curve for all hadrons (Fig.~\ref{fig:v2scaled}, right panel), indicating that the flow pattern is originally developed at the quark level.\footnote{I am indebted to A.~Bialas for the observation that the anisotropy of collective flow (elliptic flow) {\it must} be manifest at the quark level, if it is established as early in the collision ($\tau < 5$ fm/c) as hydrodynamics suggests.}

\begin{figure}[htb]
\centerline{\includegraphics[width=4.0in]{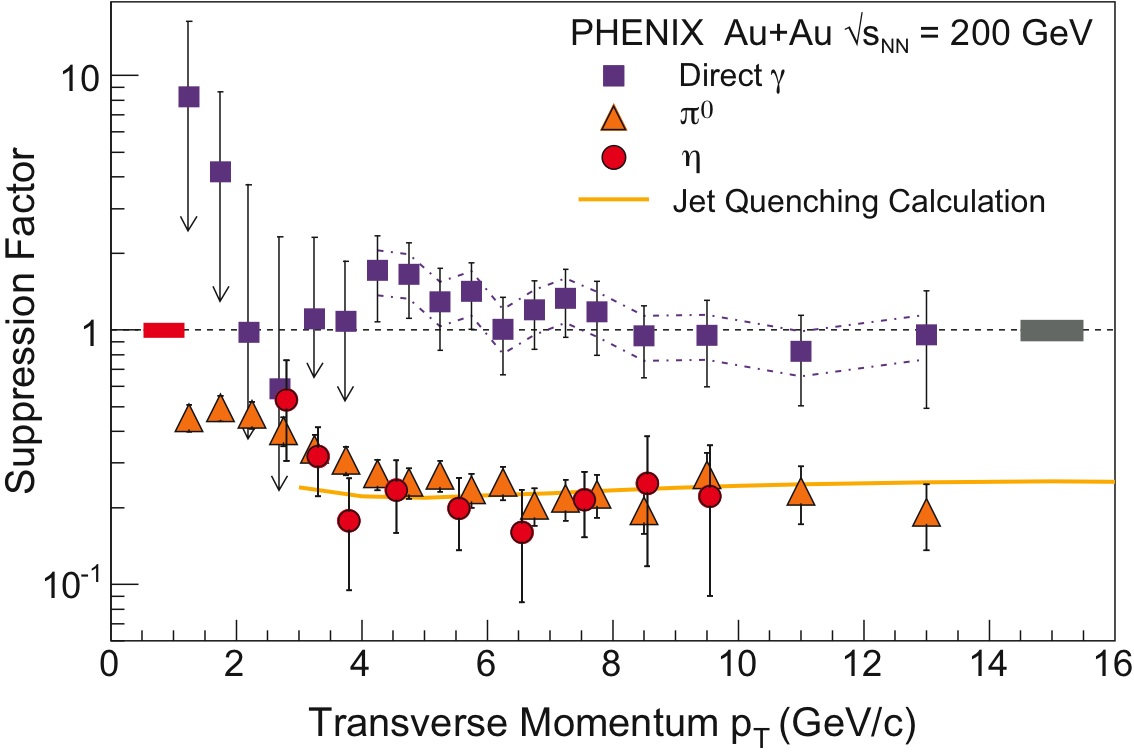}}
\caption{The ratio $R_{AA}$ of measured versus expected yield of various particles ($\pi^0, \eta, \gamma$) in Au+Au collisions at $\sqrt{s_{\rm NN}} = 200$ GeV as function of the transverse momentum $p_T$. Whereas pions and $\eta$-mesons show the same amount of suppression at high $p_T$, where quark fragmentation is the dominant production mechanism, direct photons are found to be unsuppressed. This indicates that the suppression is a final-state effect related to the absorption (energy loss) of energetic partons in the medium.}
\label{fig:RAA}
\end{figure}

{\it Jet quenching:} QCD jets arise from the hard scattering of incoming quarks and gluons and their subsequent fragmentation into directionally aligned hadrons. The rates for jet production and other hard-scattering processes grow rapidly with increasing collision energy.  The production rates of such hard probes can be accurately calculated  and their interactions with the medium can be described perturbatively. The access to hard probes was a primary motivation for constructing RHIC with high center-of-mass energy. This strategy has been validated by the discovery of jet quenching and its development as a quantitative tomographic probe of the quark-gluon plasma. Figure~\ref{fig:RAA} illustrates this discovery, showing the suppression of $\pi^0$ and $\eta$ meson emission in central Au+Au collisions compared with expectations from measurements in p+p collisions. The suppression of high-$p_T$ hadrons stands in distinct contrast to the lack of suppression seen in direct photon yields, which are consistent with perturbative QCD calculations of their initial production rate. 

In QCD, the energy loss of an highly energetic light parton (quark or gluon) is thought to be dominated by gluon radiation induced by the multiple scattering of the parton on color charges in the medium. The energy loss after passage through a homogeneous medium of length $L$ can be expressed as $\Delta E \approx - \alpha_s \hat{q} L^2$, where the stopping power the medium is governed by the parameter\cite{Baier:1996kr}
\begin{equation}
\hat{q} = \rho \int dq^2 \, q^2 \, \frac{d\sigma}{dq^2} \sim \frac{\mu^2}{\lambda_{\rm f}} .
\label{eq:qhat}
\end{equation}
Here $\rho$ denotes the density of color charges in the medium and $d\sigma/dq^2$ is the differential scattering cross section for a parton on a color charge. $\mu$ is the inverse color screening length (see eq.~(\ref{eq:mu2})) and $\lambda_{\rm f}$ denotes the mean free path of an energetic parton in the medium. The energy loss parameter $\hat{q}$ can also be expressed as a correlation function of gluon fields in the medium along the light cone \cite{Baier:1996sk,Majumder:2007hx}:
\begin{equation}
\hat{q} = \lim_{\xi\to 0} \frac{g^2 C_R}{N_c^2-1} \int dy^- \left\langle F^{ai+}(0)F_{i}^{a+}(y^-) \right\rangle 
  e^{i \xi p^+y^-} ,
\label{eq:qhatFF}
\end{equation}
where $C_R$ is the eigenvalue of the SU(3) Casimir operator for the penetrating parton.

The experimental data from RHIC can be used to determine the value of $\hat{q}$ for the quark-gluon plasma produced in nuclear collisions. This determination requires a realistic modeling of the reaction geometry and its time evolution. Such analyses are just now being performed using three-dimensional relativistic hydrodynamics \cite{Renk:2006sx,Majumder:2007ae,Qin:2007ys}. They yield values for $\hat{q}$ in the range of $0.5-20$ GeV$^2$/fm (normalized to the conditions at time $\tau=1$ fm/c), considerably larger than the original predictions made on the basis of perturbation theory. The large variation of the deduced values of $\hat{q}$ is caused by differences in treating the multiple radiation leading to energy loss in the various approaches. 

Additional evidence for the color opacity of the medium is seen in studies of the angular correlation of the radiation associated with a high-$p_T$ trigger particle \cite{Adler:2002tq,Adler:2005ee}. In p+p and d+Au collisions, a hard recoiling hadron frequently occurs at 180  degrees in azimuth to the trigger, reflecting the back-to-back nature of jets in leading order QCD. In sharp contrast, central Au+Au collisions show a strong suppression of such recoils, accompanied by an enhancement and broadening of low-$p_T$ particle production. Detailed analyses indicate that the response of the medium to the passage of an energetic parton may be of a characteristic hydrodynamical nature: the energy lost by high-energy parton may re-appear as a collective Mach cone.

\section{Toward the ``perfect'' liquid}

As stated above, a ``perfect'' liquid is a fluid with the smallest shear viscosity allowed by the laws of nature. That quantum mechanics imposes a lower limit on the shear viscosity of an entropy carrying fluid can be easily seen as follows. In kinetic transport theory, the shear viscosity $\eta$ is governed by the rate of momentum transport in the fluid:
\begin{equation}
\eta \approx \frac{1}{3} n \bar{p} \lambda_{\rm f} = \frac{\bar{p}}{3\sigma_{\rm tr}} ,
\label{eq:eta}
\end{equation}
where $n$ is the density, $\bar{p}$ the average momentum of particles in the medium, and $\sigma_{\rm tr}$ their transport cross section. In quantum mechanics, cross sections are bounded by unitarity, which poses a {\it lower} bound on the shear viscosity. E.~g., for s-wave scattering:
\begin{equation}
\sigma_{\rm tr}(p) \leq \frac{4\pi}{p^2} \qquad \longrightarrow \qquad
  \eta \geq \frac{\bar{p}^3}{12\pi} .
\label{eq:etaQM}
\end{equation}
A slightly different line of arguments rewrites (\ref{eq:eta}) in terms of the energy per particle $E/N =\varepsilon/n$ and the mean time between scatterings $\tau_{\rm f} = \lambda_{\rm f}/\bar{v}$, and then makes use of the relation $S \approx 4N$ for the entropy $S$ of  a highly relativistic system of particles. Finally, the argument recognizes that the uncertainty relation sets the bound $(E/N)\tau_{\rm f} \geq \hbar$ to obtain
\begin{equation}
\eta \approx \frac{1}{12} s (E/N) \tau_{\rm f} \geq \frac{\hbar}{12} s \approx \frac{\hbar}{4\pi} s .
\label{eq:etamin}
\end{equation}

All known materials obey this bound on $\eta$; in fact, until recently, the lowest values of $\eta$ for all materials exceeded the bound (\ref{eq:etamin}) by at least a factor ten!\footnote{This statement includes helium ($^4$He) below below the lambda point, as long as its superfluid and normal components flow together in bulk. Although the fraction of the entropy carrying normal component tends to zero as $T\to 0$, its shear viscosity increases so rapidly that, overall, $\eta/s \to \infty$.} Quite generally, the value of $\eta/s$ for any material grows at low and high temperatures and reaches a minimum somewhere in between. The minimum is usually associated with a phase transition; if the transition is of first order, a discontinuous jump in $\eta/s$ is observed. 

The only system for which the ratio $\eta/s$ is known to saturate the lower bound (\ref{eq:etamin}), at least theoretically, is thermal matter governed by the $N=4$ supersymmetric  SU($N_c$) Yang-Mills theory in the combined limit of strong coupling and large $N_c$ \cite{Policastro:2001yc}. In this limit the theory can be solved exactly by mapping it on a weakly coupled superstring theory in five-dimensional Anti-deSitter space (more precisely, the space ${\rm AdS}_5\times{S}_5$), the famous AdS/CFT duality \cite{Maldacena:1997re}. The shear viscosity is found to be (setting $\hbar = 1$) \cite{Buchel:2004di}:
\begin{equation}
\eta = \frac{s}{4\pi} \left[ 1 + \frac{135\,\zeta(3)}{(8g^2N_c)^{3/2}} +\cdots \right] .
\label{eq:etaSYM}
\end{equation}
There are also more general arguments suggesting that any system that has a gravity dual satisfies the bound (\ref{eq:etamin}) \cite{Kovtun:2004de}. Quite recently, a real system has been found in remarkable parallelism with the developments at RHIC, which comes close to the lower bound on $\eta/s$: the dilute, degenerate Fermi gas of $^6$Li atoms at sub-$\mu$K temperatures and artificially induced (via a Feshbach resonance) strong coupling among its atoms. The best experimental values for this system presently indicate that it comes within a factor of three of the $\eta/s$ bound.

\begin{figure}[htb]
\includegraphics[width=4.5in,angle=0]{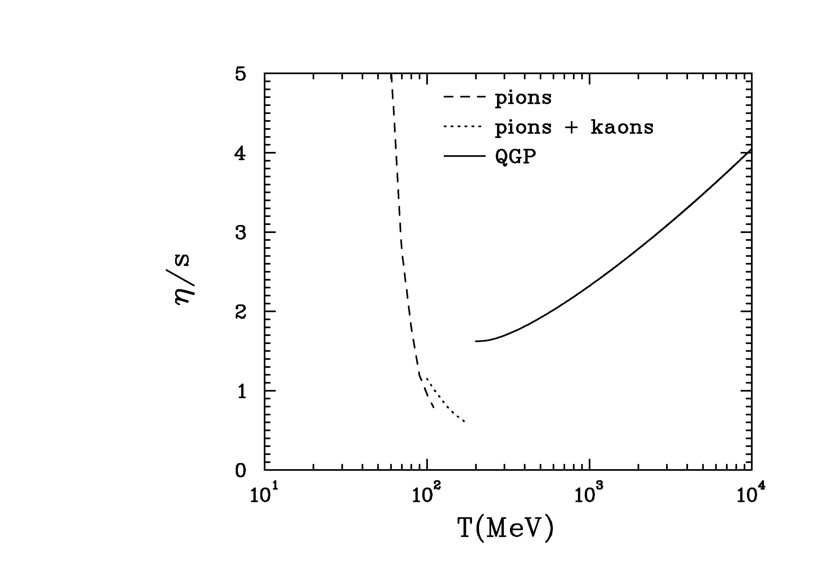}
\caption{Predicted shear viscosity-to-entropy density ratio for QCD matter as function of temperature. The solid line is the leading-order prediction for a perturbative quark-gluon plasma. The dashed line shows the result for a dilute hadronic gas using pion cross section data.}
\label{fig:eta-s}
\end{figure}

What about QCD matter? The shear viscosity of thermal QCD matter has been calculated in perturbation theory by evaluating the lowest order contributions to parton-parton scattering. At leading logarithmic order and for $N_f = 3$ the result is \cite{Arnold:2000dr}:
\begin{equation}
\eta_{\rm C} \approx \frac{9s}{100\pi\, \alpha_s^2\,\ln\alpha_s^{-1}} .
\label{eq:etaC}
\end{equation}
The index ``C'' is intended to indicate that this is the collisional shear viscosity. $\eta_{\rm C}$ has also been calculated numerically at leading order $\alpha_s^{-2}$ beyond the leading logarithm \cite{Arnold:2003zc}. When the result is plotted as a function of $T$, one finds that $\eta/s$ should exceed unity for a quark gluon plasma at all temperatures (see Fig.~\ref{fig:eta-s}) \cite{Csernai:2006zz}. The shear viscosity of a hadronic gas due to collisions among pions (and kaons) has also been calculated using, both, experimental cross section data\cite{Prakash:1993bt} and chiral perturbation theory \cite{Chen:2006iga}. The conclusion is that $\eta/s$ may fall slightly below unity at $T>100$ MeV but rises very rapidly as the temperature decreases (see Fig.~\ref{fig:eta-s}).

On the other hand, the RHIC data seem to demand a value of $\eta/s$ close to the bound (\ref{eq:etamin}). When the elliptic flow is calculated in dimensionally reduced scenarios -- only treating the transverse expansion while maintaining longitudinal boost invariance -- one finds that even a value of $\eta = s/4\pi$ seriously affects the agreement with the data obtained in ideal hydrodynamics \cite{Teaney:2003kp,Baier:2006gy,Song:2007fn}. Obviously, there is a problem. On the one hand, the calculation clearly show that the elliptic flow is generated while the matter has an energy density significantly exceeding $\varepsilon(T_c) \approx 1$ GeV/fm$^3$. On the other hand, the agreement between data and theory requires $\eta/s \ll 1$.  One concludes that, either the perturbative calculation of the shear viscosity of a quark-gluon plasma fails by a wide margin, or the shear viscosity of the matter produced in nuclear collisions is not dominated by collisions among partons. The first alternative has inspired the hypothesis that the quark-gluon plasma in the temperature range $T_c \leq T \leq 2T_c$ relevant to the nuclear collisions at RHIC is a {\it strongly coupled} plasma \cite{Gyulassy:2004zy}, nicknamed the sQGP, which is not composed of partonic quasi-particles and may even not be described by quasi-particles at all.
 
\section{Anomalous viscosity}

\begin{figure}[htb]
\vspace{0.2in}
\centerline{\includegraphics[width=2.0in]{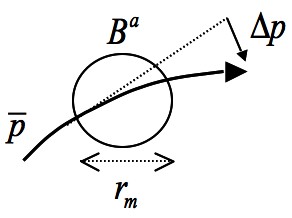}
\hspace{0.8in} \includegraphics[width=2.0in]{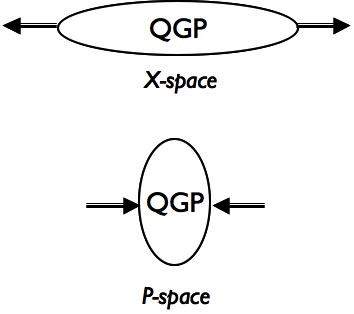}}
\caption{Left diagram: Thermal partons are deflected when they pass through a randomly oriented color field domain of sixe $t_{\rm m}$. Right diagram: The longitudinal expansion in coordinate space implies an oblate momentum distribution of partons in the medium.}
\label{fig:sketches}
\end{figure}

The second alternative does not discard the quasi-particle picture altogether, but to recall that transport processes in electromagnetic plasmas are often not dominated by collisions among particles, but by fields permeating the plasma. Since the quark-gluon plasma is a plasma, after all, it is not inconceivable that similar mechanisms are at work in the matter formed at RHIC. Indeed, we have seen earlier (Fig.~\ref{fig:domains}) that plasma instabilities generate domains of turbulent color fields whenever the parton momentum distribution is anisotropic. This will be the case before thermal equilibration is achieved, and it will remain to be true as long as the matter expands rapidly in the longitudinal direction. In fact, the momentum anisotropy in the later steady state regime is proportional to the shear viscosity. The larger the value of $\eta$, the more anisotropic is the momentum distribution of the expanding quark-gluon plasma (see Fig.~\ref{fig:sketches}, right panel).

Thus, let us assume that coherent domains of color field exist in the plasma. The orientation of the field in color space will be random from one domain to the next, but the domains will be characterized by an average field energy density and a correlation length $r_{\rm m}$. Let us, for the moment, assume that the domain is described by a color-magnetic field $B^a$. The momentum deflection of a color charge $Q^a$ passing through the domain at the speed of light is given by $|\Delta p| \approx g Q^a B^a r_{\rm m}$, as illustrated in Fig.~\ref{fig:sketches} (left diagram). Since the domains are assumed to be randomly oriented, the momentum direction of a parton will be randomized after traveling a distance
\begin{equation}
\lambda_{\rm f}^{\rm (A)} \approx \frac{\bar{p}^2}{\langle (\Delta p)^2 \rangle} r_{\rm m}
  \approx \frac{\bar{p}^2}{g^2 Q^2 \langle B^2 \rangle  r_{\rm m}} ,
\label{eq:lambdafA}
\end{equation}
where again $\bar{p}$ denotes the average thermal momentum of a parton in the plasma. Inserting this expression into the general kinetic theory formula (\ref{eq:eta}) for the shear viscosity and substituting $s \approx 4n$, $\bar{p} = 3T$, we obtain the {\it anomalous} viscosity \cite{Asakawa:2006tc}:
\begin{equation}
\eta_{\rm A} \approx \frac{9 s T^3}{4 g^2 Q^2 \langle B^2 \rangle  r_{\rm m}}  .
\label{eq:etaA}
\end{equation}
If we use the scaling estimates $\langle B^2 \rangle \sim g^2 T^4$, $r_{\rm m} \approx (gT)^{-1}$ mentioned before, we obtain $\eta_{\rm A}/s \sim g^{-3} \sim \alpha_s^{-3/2}$, which suggests that the anomalous viscosity may dominate over the collisional viscosity in the weak coupling limit.

In order to obtain a quantitative prediction, we need to calculate the anomalous viscosity in the same formal kinetic theory framework as the collisional viscosity \cite{Asakawa:2006jn}. We start from the perturbed equilibrium distribution of partons:
\begin{equation}
f(p) = f_0(p) \left[ 1 + f_1(p) (1 \pm f_0(p)) \right] ,
\label{eq:fp}
\end{equation}
where $f_0(p) = [\exp(u_\mu p^\mu/T) \mp 1]^{-1}$ is the Bose (gluons) or Fermi (quarks) distribution. The perturbation $f_1(p)$ for shear strain is parametrized by the shear viscosity $\eta$:
\begin{equation}
f_1(p) = \frac{5 \eta/s}{2 E_p T^2}\, p^i p^k \left( \nabla_i u_k + \nabla_k u_i 
  - \frac{2}{3}\delta_{ik} \nabla\cdot u \right) 
\label{eq:f1}
\end{equation}
The phase space distribution of partons, $f(p)$, is governed by the non-abelian Vlasov-Boltzmann equation
\begin{equation}
\left[ \partial_t + v\cdot\nabla_r + F\cdot\nabla_p \right] f(r,p,t) = C[f] ,
\label{eq:VBE}
\end{equation}
where $v=p/E_p$ is the parton velocity, $\vec{F} = gQ^a(\vec{E}^a + \vec{v}\times\vec{B}^a)$ the color Lorentz force, and $C[f]$ denotes the collision term. If we assume that the color fields are randomly oriented and distributed, the ensemble averaged phase space distribution $\bar{f}(p)$ can be shown to satisfy an equation of the Fokker-Planck-Boltzmann type:
\begin{equation}
\left[ \partial_t + v\cdot\nabla_r - \nabla_p\cdot D(p) \cdot\nabla_p \right] \bar{f}(r,p,t) = C[\bar{f}] ,
\label{eq:FPBE}
\end{equation}
where the momentum diffusion tensor $D_ik$ is given by
\begin{equation}
D_{ik}(p) = \int_{-\infty}^t dt' \langle F_i(r',t') F_k(r,t) \rangle \qquad {\rm with}\; r' = r+v(t'-t) .
\label{eq:Dik}
\end{equation}
For isotropically distributed color field domains with a correlation length $r_{\rm m}$ the diffusion tensor takes the form $D_{ik}(p) = \langle F^2 \rangle r_{\rm m}$. Following the standard Chapman-Enskog approach for the calculation of transport coefficients, we obtain the following result for the shear viscosity:
\begin{equation}
\frac{1}{\eta} = O(1) \frac{\langle F^2 \rangle r_{\rm m}}{(N_c^2-1) s T^3}
  + O(10^{-2})\, \frac{g^4 \ln g^{-1}}{T^3}
  \equiv \frac{1}{\eta{\rm A}} + \frac{1}{\eta{\rm C}} .
\label{eq:etatot}
\end{equation}

This expression exhibits several remarkable features. First, the numerator of the anomalous contribution to the shear viscosity, $\langle F^2 \rangle r_{\rm m}/(N_c^2-1)$, is precisely the contribution of the color field domains to the energy loss parameter $\hat{q}$. This is easily seen from the formulation of $\hat{q}$ as a gluon correlation function along the light cone, eq.~(\ref{eq:qhatFF}). Secondly, a more detailed scaling analysis reveals that the quantities $\langle F^2 \rangle$ and $r_{\rm m}$ are functions of the momentum space anisotropy of the plasma and thus, by virtue of (\ref{eq:f1}), of $\eta$ itself. Taking this implicit dependence into account, one finds that the anomalous viscosity scales as
\begin{equation}
\eta_{\rm A}/s \sim (g^2 |\nabla u|)^{-3/5} ,
\label{eq:etaAscal}
\end{equation}
where $|\nabla u|$ denotes the magnitude of the shear strain. The explicit dependence of $\eta_{\rm A}$ on the shear strain gives meaning to the term ``{\it anomalous}'' viscosity. Since the normal, collisional viscosity scales as $\eta_{\rm C} \sim (g^4 \ln g^{-1})^{-1}$, but is independent of $|\nabla u|$, the collisional viscosity always dominates for very small shear strains, whereas the anomalous viscosity dominates for large shear strains and weak coupling! More quantitative statements require the numerical evaluation of the correlator (\ref{eq:Dik}).

\section{Exploring the ``perfect'' liquid}

\begin{figure}[htb]
\centerline{\includegraphics[width=4.5in]{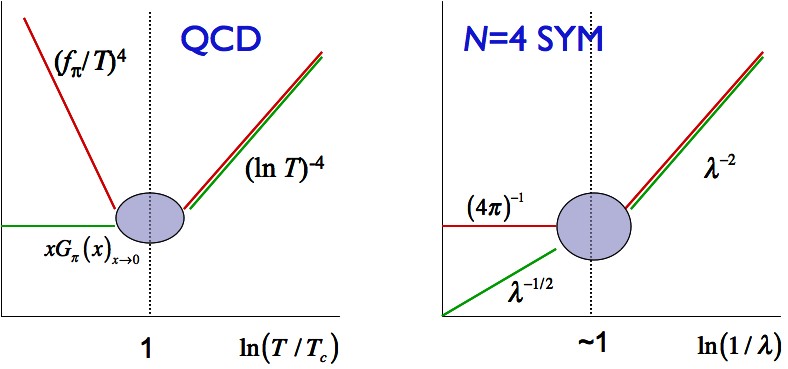}}
\caption{The left-hand side ($\eta/s$, red lines) and right-hand side ($1.25\, T^3/\hat{q}$, green lines) of eq.~(\protect\ref{eq:etaqhat}) plotted versus the effective coupling strength. Left panel: QCD; right panel: $N=4$ super-Yang-Mills theory. For QCD, the abscissa shows $T/T_c$ on a logarithmic scale; for the SYM theory, the abscissa shows the inverse 't Hooft coupling, also on a logarithmic scale.}
\label{fig:QCDSYM}
\end{figure}

It turns out that the relationship between $\hat{q}$ and $\eta$ noticed in the case of the anomalous viscosity can be generalized to any situation where the thermal excitations of the medium have the same quantum numbers and interactions as highly energetic (``hard'') excitations. It not only applies to QCD, but to any unbroken gauge theory, in which transport cross sections are dominated by small angle scattering. Under these conditions, one can derive the following relation \cite{Majumder:2007zh}:
\begin{equation}
\frac{\eta}{s} = 1.25\, \frac{T^3}{\hat{q}} .
\label{eq:etaqhat}
\end{equation}
This correspondence fails to hold when the medium is so strongly coupled that its thermal excitations cannot be described as quasi-particles, as in strongly coupled super-Yang-Mills (SYM) theory, or when the thermal excitations are described by quasi-particles of a different kind, as in QCD at temperatures below $T_c$. This is illustrated in Fig.~\ref{fig:QCDSYM}, where the left-hand side and the right-hand side of eq.~(\ref{eq:etaqhat}) are plotted separately versus the temperature (for QCD) or the 't Hooft coupling $\lambda = g^2N_c$ (for the SYM theory). In both cases, the two sides of the equation diverge when the medium is not described by a partonic quasi-particle perturbation theory. In both instances, the right-hand side of (\ref{eq:etaqhat}) is a better measure of the fundamental coupling strength when the relationship fails to hold. 

Reliable independent determinations of $\eta/s$ and $\hat{q}/T^3$ from the RHIC data would thus permit a model independent determination of the strongly coupled nature of the matter produced in the collisions. If the relation (\ref{eq:etaqhat}) holds, the quark-gluon plasma at RHIC is composed of partonic quasi-particles; if it fails, it is a strongly coupled quark-gluon plasma (sQGP). Presently, the analysis of the RHIC data in terms of $\eta/s$ and $\hat{q}/T^3$ is not yet precise enough to make this determination possible. Using the value $\hat{q}_0 \approx 1-2$ GeV$^2$/fm extracted in the higher-twist analysis of jet quenching\cite{Zhang:2007ja} and the initial temperature $T_0 \approx 335$ MeV deduced from the measured entropy of the final state\cite{Muller:2005en}, one finds $1.25\, T^3/\hat{q} \approx 0.12 - 0.24$, not inconsistent with the constraints on $\eta/s$ obtained from comparisons of viscous hydrodynamics calculations with the elliptic flow data.

\begin{figure}[htb]
\centerline{\includegraphics[width=4.5in]{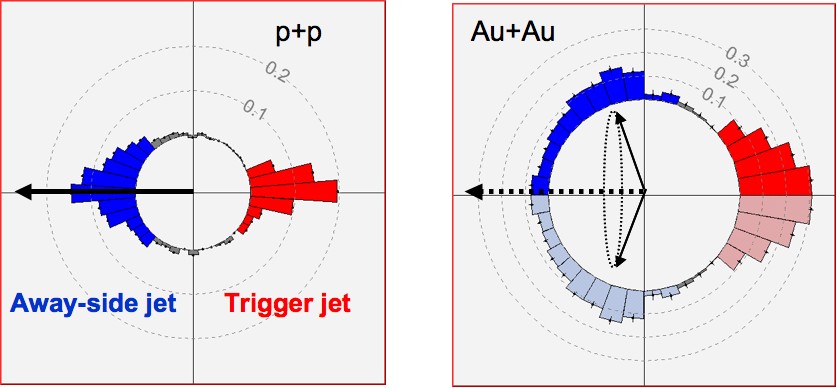}}
\caption{Angular distribution of secondaries around the beam axis with respect to the direction of the trigger hadron. The red part shows the jet accompanying the triggered hadron, the blue part shows the fragments of the recoil parton. Left panel: p+p collisions exhibit the usual $180^\circ$ di-jet event structure. Right panel: In Au+Au collisions the peak of the recoil secondaries is shifted to $110^\circ$ in azimuth, suggestive of a Mach cone around the recoil direction.}
\label{fig:Mach}
\end{figure}

If the effect of the medium on a hard parton can tell us something about the properties of the matter formed at RHIC, then the fate of the energy lost by the parton must be even more informative. This issue has two aspects: the energy lost by the triggered parton and the energy lost by its recoil partner. The latter is far larger than the former because, on average, the triggered parton originates close to the near-side surface of the fireball and thus traverses only a relatively small amount of matter, while the recoil partner, again on average, propagates through the bulk of the fireball. In fact, the recoil parton loses so much of its energy that its fragments become virtually indistinguishable from the matter itself. However, the energy does not disappear, and its kinematic distribution can be identified by careful background subtraction. The surprising result of this analysis was that the excess energy does not appear at $180^\circ$ in azimuth with respect to the trigger, but at an angle of about $110^\circ$ (see Fig.~\ref{fig:Mach}. A three-body coincidence analysis indicates that the energy appears on both sides of the recoil, ruling out a simple deflection mechanism.  

The explanation most compatible with the data, but also the most speculative and spectacular one, is that the energy lost by the recoil parton travels through the medium in the form of a sonic Mach cone \cite{CasalderreySolana:2004qm}. If this is true, the angular peak position can be used to determine the sound velocity of the quark-gluon plasma! It would also show that sound propagation in the medium is only weakly damped and thus confirm that the shear viscosity is small. The formation of a Mach cone trailing a fast color charge has been demonstrated by explicit calculations in the strongly coupled SYM theory \cite{Friess:2006fk}. In QCD, the precise form of the coupling between a fast color charge and the sound mode has not yet been derived. What is clear from phenomenological studies is that the coupling must be highly (about 70\%) effective in terms of energy transfer \cite{Renk:2005si}.

\begin{figure}[htb]
\vspace{0.2in}
\centerline{\includegraphics[width=2.2in]{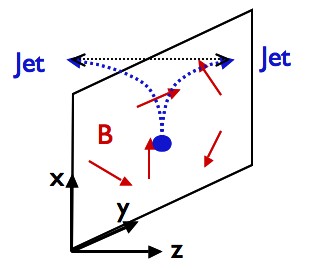}
\hspace{0.4in} \includegraphics[width=2.2in]{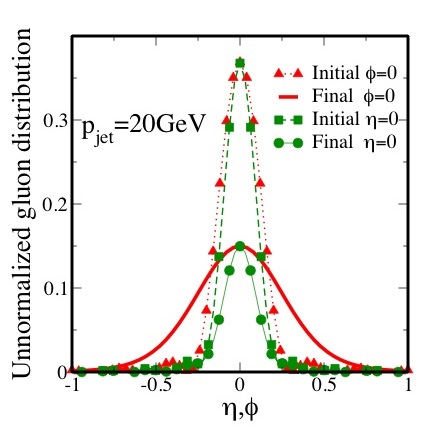}}
\caption{Left panel: Schematic representation of the longitudinal deflection of radiated gluons propagating through transversely oriented color-magnetic fields. Right panel: Longitudinal broadening of the distribution of radiated gluons obtained by solution of the diffusive Boltzmann equation (\protect\ref{eq:FPBE}). The final pseudorapidity distribution is shown by the solid red curve.}
\label{fig:ridge}
\end{figure}

The emission pattern of secondaries on the trigger side is also modified in Au+Au collisions compared with p+p collisions. In addition to a jet-like pattern of secondaries at small angles with respect to the trigger hadron with nearly the same characteristics as in p+p collisions, one finds also a wide ridge near $0^\circ$ in azimuth but distributed over a wide pseudorapidity range ($\Delta\eta > 3$) \cite{Putschke:2007mi}. The chemical composition and spectral distribution of the hadrons making up the ridge is very similar to the bulk matter. One possible explanation is that the ridge is formed by the energy lost by the parton that produced the trigger hadron. Since the jet cone itself has the same shape in p+p and Au+Au, it must be formed in vacuum,  after the parton escapes from the fireball. But why does the energy radiated during its passage through the matter spread out longitudinally, but not in azimuth? This could actually be the result of the deflection of the radiated gluons by turbulent color fields present in the quark-gluon plasma. If these are mostly color-magnetic fields oriented in the transverse direction to the beam axis, as it is suggested by a perturbative analysis of the color field instabilities, they would result in a preferentially longitudinal deflection of the radiated gluons (see Fig.~\ref{fig:ridge}, left panel). The diffusion in pseudorapidity $\eta$, but not in azimuthal angle $\phi$ is, indeed, found when one solves the diffusive transport equation (\ref{eq:FPBE}) for radiated gluons in the presence of transversely oriented color-magnetic fields (see Fig.~\ref{fig:ridge}, right panel) \cite{Majumder:2006wi}. Similar effects have recently also been found in self-consistent microscopic simulations of partons propagating in spontaneously generated color fields \cite{Dumitru:2007rp}.

\section{Conclusion}

The incredible wealth and quality of the RHIC data have challenged theoretical ideas of the physical properties of the quark-gluon plasma on many fronts. The most remarkable insight gleaned from the data is that the matter produced in nuclear collisions at RHIC is an almost `` perfect'', i.~e.\ nearly inviscid liquid, whose shear viscosity approaches the lower bound dictated by unitarity. The strongly coupled super-Yang-Mills theory provides one model for such a liquid; an alternative model is the turbulent plasma permeated by color fields generated by expansion driven plasma instabilities. If both, shear viscosity $\eta$ and energy loss parameter $\hat{q}$ of the matter can be determined reliably from an analysis of the data, it may be possible to distinguish between the two scenarios. The found relation, connecting a large color opacity of the quark-gluon plasma with a low dissipation for collective flow, is clearly confirmed by the RHIC data. Increasingly differential studies of the phenomena accompanying jet formation in nuclear collisions are promising further surprises, which will keep theorists occupied, if not puzzled, for years to come. 

\section*{Acknowledgements}

I thank the organizers of the Zakopane School of Theoretical Physics for the invitation to lecture at this school and the participating students for their enthusiasm and interest. This work was supported, in part, by a grant from the U. S. Department of Energy (DE-FG02-05ER41367).


\end{document}